\documentclass{ws-procs9x6}

\usepackage{latexsym}

\newcounter{subequation}[equation]
\makeatletter

\def\bcite{\@ifnextchar [{\@tempswatrue\@bcitex}{\@tempswafalse\@bcitex[]}}
\def\@bcitex[#1]#2{\if@filesw\immediate\write\@auxout{\string\citation{#2}}\fi
  \let\@bcitea\@empty
  \@bcite{\@for\@bciteb:=#2\do
    {\@bcitea\def\@bcitea{,\penalty\@m\ }%
     \def\@tempa##1##2\@nil{\edef\@bciteb{\if##1\space##2\else##1##2\fi}}%
     \expandafter\@tempa\@bciteb\@nil
     \@ifundefined{b@\@bciteb}{{\reset@font\bf ?}\@warning
       {Citation `\@bciteb' on page \thepage \space undefined}}%
     \hbox{\csname b@\@bciteb\endcsname}}}{#1}}
\def\@bcite#1#2{{#1\if@tempswa , #2\fi}}

\def\thesubequation{\theequation\@alph\c@subequation}
\def\@subeqnnum{{\rm (\thesubequation)}}
\def\slabel#1{\@bsphack\if@filesw {\let\thepage\relax
   \xdef\@gtempa{\write\@auxout{\string
      \newlabel{#1}{{\thesubequation}{\thepage}}}}}\@gtempa
   \if@nobreak \ifvmode\nobreak\fi\fi\fi\@esphack}
\def\subeqnarray{\stepcounter{equation}
\let\@currentlabel=\theequation\global\c@subequation\@ne
\global\@eqnswtrue
\global\@eqcnt\z@\tabskip\@centering\let\\=\@subeqncr
$$\halign to \displaywidth\bgroup\@eqnsel\hskip\@centering
  $\displaystyle\tabskip\z@{##}$&\global\@eqcnt\@ne
  \hskip 2\arraycolsep \hfil${##}$\hfil
  &\global\@eqcnt\tw@ \hskip 2\arraycolsep
  $\displaystyle\tabskip\z@{##}$\hfil
   \tabskip\@centering&\llap{##}\tabskip\z@\cr}
\def\endsubeqnarray{\@@subeqncr\egroup
                     $$\global\@ignoretrue}
\def\@subeqncr{{\ifnum0=`}\fi\@ifstar{\global\@eqpen\@M
    \@ysubeqncr}{\global\@eqpen\interdisplaylinepenalty \@ysubeqncr}}
\def\@ysubeqncr{\@ifnextchar [{\@xsubeqncr}{\@xsubeqncr[\z@]}}
\def\@xsubeqncr[#1]{\ifnum0=`{\fi}\@@subeqncr
   \noalign{\penalty\@eqpen\vskip\jot\vskip #1\relax}}
\def\@@subeqncr{\let\@tempa\relax
    \ifcase\@eqcnt \def\@tempa{& & &}\or \def\@tempa{& &}
      \else \def\@tempa{&}\fi
     \@tempa \if@eqnsw\@subeqnnum\refstepcounter{subequation}\fi
     \global\@eqnswtrue\global\@eqcnt\z@\cr}
\let\@ssubeqncr=\@subeqncr
\@namedef{subeqnarray*}{\def\@subeqncr{\nonumber\@ssubeqncr}\subeqnarray}
\@namedef{endsubeqnarray*}{\global\advance\c@equation\m@ne%
                           \nonumber\endsubeqnarray}

\makeatother

\DeclareFontFamily{OT1}{rsfs10}{}
\DeclareFontShape{OT1}{rsfs10}{m}{n}{ <-> rsfs10 }{}
\DeclareMathAlphabet{\mathscript}{OT1}{rsfs10}{m}{n}

\newcommand{\bsea}{\begin{subeqnarray}} 
\newcommand{\esea}{\end{subeqnarray}}

\def\am{*Center For Theoretical Physics, Department of Physics\\ Texas A\&M University,
 College Station,TX 77843-4242,
USA}
\def\regina{**Department of Physics, University of Regina\\ Regina SK, S4S OA2, Canada}
\def\address#1{\begin{center}{ \it #1} \end{center}}
\def\author#1{\begin{center}{ \sc #1} \end{center}}
\def\title#1{\begin{center} {\Large #1 } \end{center}}
\def\Journal#1#2#3#4{{#1} {\bf #2}, #3 (#4)}

\def\JHEP{\em JHEP}

\def\NPB{{\em Nucl. Phys.} B}

\def\PRL{\em Phys. Rev. Lett.}
\def\PRD{{\em Phys. Rev.} D}

\begin{document}

\title{Horava-Witten Cosmology}
\author{*R. Arnowitt, *James Dent, and B. Dutta**}
\address{\am}
\address{\regina}


\abstracts{We consider the cosmology of the reduced 5D Horava-Witten M-Theory (HW) with volume modulus and treating matter on the orbifold planes to first order.  It is seen that one can recover the FRW cosmology in the Hubble expansion era with relativistic matter, but if a solution exist with non-relativistic (massive) matter it must be non-static with a Hubble constant that depends on the fifth dimension. (The same result holds when 5-branes are present.)  This difficulty is traced to the fact that in HW, the volume modulus couples to the bulk and brane cosmological constants (so that the net 4D constant vanishes naturally).  This situation is contrasted with the Randall-Sundrum 1 model (which is here treated without making the stiff potential approximation) where the radion field does not couple to the cosmological constants (and so one must instead fine tune the net constant to zero).  One finds that non-relativistic matter is accommodated there by changing the distance between the end branes.}
\section{Introduction}
The Horava-Witten M-Theory [1-4] is a natural model one might use to try to build a more fundamental theory of cosmology.  The model examines 11 dimensional (11D) supergravity on an orbifold $M_{10}$$\times$$S^{1}$/$Z_{2}$ where $Z_{2}$ is a reflection in the 11th dimension.  Alternately one may think of this space as an 11D bulk space bounded by two 10D orbifold planes at $x^{11}$=0 and $x^{11}$ = $\pi$$\rho$, where by convention physical space is $x^{11}$ = 0.

The construction of the model involves imposing the interlocking constraints of anomaly cancelation, local supersymmetry and Yang-Mills (YM) gauge invariance, and leads to a phenomenologically satisfying framework.  Thus one finds that $E_8$ Yang-Mills multiplets must exist on each orbifold plane (which can easily be broken to the Standard Model (SM) gauge group on the physical plane), three generation models exist, and the 11D Planck mass, $\kappa$, is related to the 10D YM coupling constant, $\lambda$, by
\begin{equation}
\lambda^2 = 2\pi\left(4\pi\kappa^2\right)^{2/3}
\end{equation}
This leads in the simplest case of the manifold $M_4$$\times$$X$$\times$$S^1$/$Z_2$ to [3]
\begin{equation}
G_{N}= \frac{\kappa^2}{16\pi^{2}\widetilde{V}\rho}\,\,\,\,\, ; \,\,\,\,\, \alpha_G = \frac{\left(2\pi\kappa^2\right)^{2/3}}{2\widetilde{V}} 
\end{equation}
where $\widetilde{V}$ is the CY volume of $X$ and $\alpha_G$ is the GUT scale coupling constant.  Hence for $\widetilde{V}$ = $M_{G}^{-6}$ where $M_G$ = 3$\times$$10^{16}$ GeV is the GUT mass and $\alpha_G$ = 1/24, one finds $\kappa^{-2/9}$ $\cong$  $M_G$ and 1/$\pi$$\rho$ $\cong$ 4.7$\times$$10^{15}$.  Thus one has a natural explanation of the Planck scale - GUT scale hierarchy: the fundamental parameter is the 11D $\kappa$ which sets the GUT scale while the 4D Planck mass is a derived quantity which is accidentally large due to 4$\pi$ factors.  Further, the orbifold length is $\mathnormal{O}$(10) times the CY length.

\section{The 5D Model}
The last point shows that as one goes below the GUT scale, one first encounters a 5D theory on the manifold $M_4$$\times$$S^1$/$Z_2$.  We parameterise this with coordinates $x^{\mu}$ and y = $x^{11}$, where 0 $\leq$ y $\leq$ $\pi$$\rho$.  The 5D theory includes the following moduli: $\mathnormal{V}$ = $e^{\phi}$ (the CY volume modulus), $b_i$ (the shape moduli), as well as the matter fields $F_{\mu\nu}$ (the YM field strength), $C^p$ (the chiral matter fields) and the matter superpotential W.  In addition there are topological parameters
\begin{equation}
\alpha_i = \frac{\pi}{\sqrt{2}}\left(\frac{\kappa}{4\pi}\right)^{2/3}\frac{1}{\widetilde{V}^{2/3}}\beta_i\,\,\, , \,\,\, \beta_i = -\frac{1}{8\pi^2}\int_{C_i}\mbox{tr}R\wedge R
\end{equation}
where $\beta_i$ are integers (the first Pontrjagin class of the CY).

We consider the simplest model keeping only the volume modulus V and neglecting the shape moduli and other moduli.  The action reads\begin{eqnarray}\nonumber
S & = &-\frac{1}{2\kappa_{5}^2}\int_{M_5}\sqrt{g}[R
+\frac{1}{2}V^{-2}\partial_{\alpha}V\partial^{\alpha}V+\frac{3}{2}\alpha^{2}V^{-2}]\\\nonumber& &+\frac{1}{\kappa_{5}^2}\sum_{i}\int_{M_{4}^{(i)}}\sqrt{-g}V^{-1}3(-1)^{i+1}\alpha
\\& &-\frac{1}{16\pi\alpha_{GUT}}\sum_{i}\int_{M_{4}^{(i)}}\sqrt{-g}VtrF_{\mu\nu}^{i^{2}}
\\\nonumber& &
-\sum_{i}\int_{M_{4}^{(i)}}\sqrt{-g}\bigg[(D_{\mu}C)^{n}(D_{\mu}\bar{C})^{n}+V^{-1}\frac{\partial{W}}{\partial{C^{n}}}\frac{\partial{\bar{W}}}{\partial{\bar{C^{n}}}}
+D^{(\mu)}D^{(\mu)}\bigg]
\end{eqnarray}
We note the appearance of the cosmological constant scaled by $\alpha$$V^{-1}$, both in the bulk and on the branes with coefficients predicted by the theory.  These coefficients (arising from anomaly cancellation) arrange the net cosmological constant seen on the branes to be zero without any fine tuning.

For the metric we chose the cosmological ansatz 
\begin{equation}
ds^{2}=a(t,y)^{2}dx^{k}dx^{k}-n(t,y)^{2}dt^{2}+b(t,y)^{2}dy^{2},
\end{equation}
which must obey the following orbifold boundary conditions at $y_1$ = 0 and $y_2$ = $\pi$$\rho$:
\begin{eqnarray}\nonumber
(-1)^{i}\frac{1}{b}\frac{a'}{a} \,\bigg|_{y=y_{i}}&=& \frac{\rho_{i}}{6M_{5}^{3}}\,\,\,;\,\,\,\rho_i = \rho_{ir} e^{\phi_i} + \rho_{inr} e^{-\phi_i}+ 3
  M_5^3\alpha_i e^{-\phi_i}\\\nonumber
(-1)^{i}\frac{1}{b}\frac{n'}{n} \,\bigg|_{y=y_{i}}&=& -\frac{2\rho_{i}+3p_{i}}{6M_{5}^{3}}\,\,\,;\,\,\,p_i = p_{ir} e^{\phi_i} - 3 M_5^3\alpha_i e^{-\phi_i}\\
\phi'\,\bigg|_{y=y_{i}}&=&\left[\left(3b\alpha_i - \frac{b\rho_{inr}}{M_{5}^3}\right)e^{-\phi}\right]_{y=y_i}
\end{eqnarray}
where $\rho_{ir,nr}$ and $p_{ir}$ are the energy densities and pressues on the orbifold planes. (The subscripts r, nr refer to relativistic (massless) and non-relativistic (massive) matter and $M_5$ is the 5D Planck mass.)
\section{Hubble Expansion Era}
We look for solutions of the field equations that describe the Hubble expansion era.  Here the matter density is much less than the characteristic energy scale of the Calabi-Yau manifold or the orbifold scale i.e. $\rho_{matter}$$\ll$$M_{G}^{4}$.  We can thus think of brane matter as a perturbation on the vacuum.  The vacuum solution of the field equations and boundary conditions for the metric of Eq.(5) has been obtained in [5] to be a(y) = n(y) = $f^{1/2}$; b(y) = $b_o$$f^2$; V(y) = $b_o$$f^3$, where f(y) = $c_o$ + $\alpha$$\mid$y$\mid$.  This solution preserves Poincaire invariance and breaks 4 of the 8 supersymmetries, which is appropriate for getting N=1 supersymmetry when one descends to four dimensions.  The parameters $b_o$ and $c_o$ are arbitrary due to flat directions in the potential and we take them here to be of $\mathnormal{O}$(1).  We treat matter as a perturbation
\begin{eqnarray}\nonumber
a(y,t) &= &f^{1/2}(1+\delta a(y,t))\,\,\,n(y) = f^{1/2}(1+\delta n(y))\\
b(y) &= &b_{0}f^{2}(1+\delta b(y,t))\,\,\,V(y) = b_{0}f^{3}(1+\delta V(y,t))
\end{eqnarray}
It is useful to introduce the following variables:
\begin{eqnarray}\nonumber
\Delta a' \equiv \delta a' + \frac{\alpha}{2f}\delta V - \frac{\alpha}{2f}\delta b\\\nonumber
\Delta n' \equiv \delta n' + \frac{\alpha}{2f}\delta V - \frac{\alpha}{2f}\delta b\\
\Delta V' \equiv \delta V' + \frac{3\alpha}{f}\delta V - \frac{3\alpha}{f}\delta b
\end{eqnarray}
and define the Hubble constant as H = $\dot{a}$/a.  Here prime means y derivative and dot denotes time derivative.

The significance of the variables of Eq.(8) is that they are invariant under a y coordinate re-parameterization $\bar{y}$ = $y$ + f($y$) where f(y) is $\mathnormal{O}$($\delta$a).  Thus these are the natural variables that will appear in the field equations.  To linear order, the field equations read:
\begin{eqnarray}
\Delta a'' = b_{o}^2 f^3\left(H^2+H\frac{\dot{b}}{b} - \frac{1}{12}\dot{\phi}^2\right)\equiv b_{o}^2
f^3A_1 \\
\Delta n'' +2\Delta a'' = b_{o}^2 f^3\left(3H^2+2\dot{H}+2H\frac{\dot{b}}{b} + \frac{\ddot{b}}{b} + 
\frac{\dot{\phi}^2}{4}\right)\equiv b_{o}^2 f^3A_2\\
3\Delta a' +\Delta n' -\Delta V' = \frac{b_{o}^2 f^4}{\alpha}\left(4H^2+2\dot{H} -
\frac{\dot{\phi}^2}{6}\right)\equiv \frac{b_{o}^2 f^4}{\alpha}A_3\\
\Delta V'' +\frac{3\alpha}{f}\left(\Delta n' +3\Delta a' - \Delta V' \right)
=b_{o}^2f^3\left(\ddot{\phi}+3H\dot{\phi}+\frac{\dot{b}}{b}\dot{\phi}\right)
\equiv b_{o}^2f^3A_4
\end{eqnarray}
The boundary conditions at y = $y_i$ of Eq.(6) to linear order are
\begin{eqnarray}
\Delta a_i' = (-1)^i\left(\frac{b_{o}^2f_{i}^5}{6M_{5}^{3}}\rho_{ir}+\frac{1}{6M_{5}^{3}f_i}\rho_{inr}\right)\\
\Delta n_i' = (-1)^{i+1}\left(\frac{b_{o}^2f_{i}^5}{6M_{5}^{3}}\left(2\rho_{ir}+3p_{ir}\right)+\frac{1}{6M_{5}^{3}f_i}2\rho_{inr}\right)\\
\Delta V_i' = (-1)^{i+1}\frac{1}{M_{5}^{3}f_i}\rho_{inr}
\end{eqnarray}
and $f_i$ = $f(y_i)$.  It is important to note that the $\emph{same}$ variables, $\Delta$a$'$, $\Delta$n$'$, $\Delta$V$'$ that appear in the field equations also enter into the boundary conditions on the orbifold planes.  While coordinate invariance requires the use of these variables in the bulk, it is not necessary that they also are the quantities that occur at the boundaries, and the fact that they do is a special feature of the Horava-Witten model.  Thus the $\Delta$$V_{i}'$ combination of Eq.(15) arises from the brane cosmological constant term in Eq.(4) (which as mentioned above is just what is needed to cancel the total cosmological constant on the brane).  The fact that $\Delta$$V_{i}'$ appears in Eq.(15) has a special significance that we will discuss in detail below.

It is easy to solve the field equations and impose the boundary conditions.  One determines in this way $\Delta$a$'$, $\Delta$n$'$, $\Delta$V$'$ plus an additional integral constraint from the $G_{yy}$ equation.  The results are
\begin{subeqnarray}
\Delta a' = \frac{b_{o}^2}{\alpha}\left( \int_{f_1}^{f}df'f'^{3} A_1 + c_1\right)\\
c_1 = -\frac{\lambda}{6}\left(b_{o}^2f_{1}^5\rho_{1r}+\frac{1}{f_1}\rho_{1nr}\right)\,\,\,;\,\,\, \lambda = \frac{\alpha}{b_{o}^{2}M_{5}^3}\\
\int_{f_1}^{f_2}df'f'^3A_1 = \frac{\lambda}{6}\left(b_{o}^2\left(\rho_{1r}f_{1}^5+\rho_{2r}f_{2}^5\right) + \left(\frac{\rho_{1nr}}{f_1}+\frac{\rho_{2nr}}{f_2}\right)\right)
\end{subeqnarray}\vspace{-.5cm}
\begin{subeqnarray}
3\Delta a' + \Delta n' = \frac{b_{o}^2}{\alpha}\left( \int_{f_1}^{f}df'f'^{3}\left(A_1+A_2\right) + c_1 + c_2\right)\\
c_1 + c_2 = -\frac{\lambda}{6f_1}\rho_{1nr}\\
\int_{f_1}^{f_2}df'f'^{3}\left(A_1+A_2\right) =
\frac{\lambda}{6}\left(\frac{1}{f_1}\rho_{1nr}+\frac{1}{f_2}\rho_{2nr}\right)
\end{subeqnarray}\vspace{-.5cm}
\begin{subeqnarray}
\Delta V' = \frac{b_{o}^2}{\alpha}\left( \int_{f_1}^{f}df'f'^{3}\left(A_4-3A_3\right) + c_3\right)\,\,\,;\,\,\,c_3 =
\frac{\lambda}{f_1}\rho_{1nr}\\
\int_{f_1}^{f_2}df'f'^{3}\left(3A_3-A_4\right) = \lambda\left(\frac{1}{f_1}\rho_{1nr}+\frac{1}{f_2}\rho_{2nr}\right)
\end{subeqnarray}\vspace{-.5cm}
\begin{equation}
\frac{b_{o}^2}{\alpha}\left(\int^{f}_{f_1}df'f'^3\left(A_1+A_2+3A_3-A_4\right) +c_1+c_2-c_3\right) = \frac{b_{o}^2f^4}{\alpha}A_3
\end{equation}

We consider first the static case where $\dot{\phi}$ = 0 = $\dot{b}$.  Here using the definitions of $A_i$ in Eqs.(9-12) one has that $A_4$ = 0 and $A_3$ = $A_1$ + $A_2$ = 4$H^2$ + 2$\dot{H}$.  Hence 3$\times$ Eq.(17c) - Eq.(18b) yields the constraint:
\begin{equation}
\frac{\lambda}{2}\sum_{i}\rho_{inr} = 0.
\end{equation}
Thus the HW equations do not tolerate a static metric with non-relativistic matter (without fine tuning between the two orbifold planes).  If we set $\rho_{inr}$ to zero, the remaining equations are totally consistent for arbitrary amounts of $\rho_{ir}$ and reproduce FRW cosmology for the radiation dominated era.  Further, the solution then is stable in the sense that if we add an additional amount of $\rho_{ir}$, we get another static solution of the same type\footnote{This result was first presented in a different form at the First International Conference on String Phenomenology (Oxford, July 2002) [6].  It is also implicit in the analysis of [7], though the authors of [7] seem not to have observed the above results}.

Horava-Witten M-Theory also allows the introduction of 5-branes in the bulk transverse to the orbifold direction, and one may generalize the above result for this case.  Again one finds that for static solutions, $\rho_{inr}$ must vanish (assuming no fine tuning between the non-relativistic energy densities on the orbifold planes and the 5-branes).

\section{Non-relativistic Matter and Time Dependence}
As shown above, Horava-Witten theory does not tolerate a static solution with massive non-relativistic matter.  We show now that if non-relativistic matter is present the only possible consistent solution (if one exists) requires either the fifth dimension length or the volume modulus (or both) to be time dependent, and also the Hubble constant must be y-dependent.  This follows directly from the $G_{yy}$ field equation, Eq.(19), which may be equivalently written as the bulk equation
\begin{equation}
A_1 + A_2 - A_3 - A_4 = f\frac{dA_3}{df}
\end{equation}
with boundary condition
\begin{equation}
A_3(y_1) = -\frac{7}{6}\lambda\frac{\rho_{1nr}}{f_{1}^{5}}
\end{equation}
Further, subtracting Eq.(17c) + Eq.(18b) from Eq.(19) evaluated at y = $y_2$ one finds
\begin{equation}
A_3(y_2) = +\frac{7}{6}\lambda\frac{\rho_{2nr}}{f_{2}^{5}}
\end{equation}
and so $A_3$ is y dependent.  But from Eq.(11), to linearized order, we have that $A_3$ = 4$H^2$ + 2$\dot{H}$.  Therefore the Hubble constant is also y dependent, i.e. H = H(y,t).  Finally, inserting the $A_i$ into Eq.(21) gives
\begin{equation}
3H\left(\frac{\dot{b}}{b}-\dot{\phi}\right) + \frac{\ddot{b}}{b} - \ddot{\phi} = f\frac{dA_3}{df}
\end{equation}
and since we have seen $A_3$ is y dependent, one must have
\begin{equation}
\frac{\dot{b}}{b} - \dot{\phi} \neq 0.
\end{equation}
Thus if a consistent solution exists with $\rho_{nr}$ present, it must be non-static and y-dependent.
\section{Randall-Sundrum (RS1) Model}
The Randall-Sundrum (RS1) model is a 5D phenomenology very similar in structure to the reduced 5D Horava-Witten theory.  RS1 has been discussed by a large number of people (see e.g. [8-11]) and we briefly compare it here to HW.  The RS action is given by
\begin{eqnarray}\nonumber
S &=& \int d^{5}x\sqrt{-g}\left(-\frac{1}{2\kappa^2}R - \Lambda + \frac{1}{2}\partial_{\mu}\phi\partial^{\mu}\phi - V\left(\phi\right)\right)\\ &+& \sum_{i=1,2}\int d^{4}x\sqrt{-g}\left(\mathcal{L}_{mi} - V_{i}\left(\phi\right)\right)
\end{eqnarray}
Here $\Lambda$ is the bulk cosmological constant, and the scalar field $\phi$ plays the role of the volume modulus.  However, note that $\phi$ does not couple to the matter (which is characterized by $\mathcal{L}_{mi}$).  The cosmological metric is chosen as [11]
\begin{equation}
ds^2 = e^{-2N}dt^2 - a_{o}(t)e^{-2A}dx^{k}dx^k - b^{2}dt^2
\end{equation}
with an expansion around the vacuum solution of A = $A_o$(y) + $\delta$A, N = $A_o$(y) + $\delta$N, b = $b_o$ + $\delta$b, and $\phi$ = $\phi_o$ + $\delta$$\phi$.  The variables that are invariant under y reparameterization now read
\begin{eqnarray}
\Delta A' & \equiv & \delta A' -A_{o}'\frac{\delta b}{b_o} -\frac{A_{o}''}{\phi_{o}'}\delta\Phi\ \\
\Delta N' & \equiv & \delta N' -A_{o}'\frac{\delta b}{b_o} -\frac{A_{o}''}{\phi_{o}'}\delta\Phi\ \\
\Delta \Phi' & \equiv & \delta\Phi' - \Phi_{o}'\frac{\delta b}{b_o} -\frac{\Phi_{o}''}{\Phi_{o}'}\delta\Phi
\end{eqnarray}
and indeed these are the quantities that appear in the bulk field equations.  The boundary conditions are (in RS the physical brane is at y=$y_2$)
\begin{eqnarray}
\Delta A_{i}' = \left(-1\right)^{i+1} \frac{\kappa^2}{6}b_o\rho_i\\
3\Delta A_{i}' + \Delta N_{i}' = \left(-1\right)^{i+1} \frac{\kappa^2}{6}\rho_{inr}\\
\Delta \phi_{i}' = \left(-\frac{\phi_{o}''}{\phi_{o}'} + \left(-1\right)^{i+1}\frac{b_o}{2}V_{i}''\left(\phi_o\right)\right)\delta\phi(y_i)
\end{eqnarray}
where $V_{i}''$ $\equiv$ $\partial^2$$V_i$/$\partial$$\phi_{i}^2$.  The important difference between RS1 and HW is that the $\phi$ boundary condition now depends $\emph{both}$ on $\Delta$$\phi_{i}'$ and $\delta$$\phi(y_i)$.  The reason for this is that in the RS phenomenology the $\phi$ field does not couple to the cosmological constants on the branes, while in HW it couples to both bulk and brane cosmological constants (and also to other matter).  It was this coupling that eliminated the $\delta$$\phi(y_i)$ term in Eq.(15) and was needed to naturally cancel the net brane cosmological constant (which, in contrast is fine tuned to zero in RS1). 

We summarize here the results obtained by solving the field equations and imposing the boundary conditions.
\begin{equation}
\delta\phi(y_i) = -\frac{(-1)^{i+1}\frac{1}{2}b_{o}\rho_{inr} + \frac{3}{\kappa^2}b_{o}^{2}\left(2H^2 + \dot{H}\right)}{\phi_{o}'' + (-1)^{i+1}\frac{b_o}{2}\phi_{o}V_{i}''}\,\bigg|_{y=y_{i}}
\end{equation}
Note that in the stiff potential limit, $V_{i}''$ $\rightarrow$ $\infty$, one has $\delta$$\phi$($y_i$) $\rightarrow$ 0 (in accord with [11]).  Eq.(34) is a generalization of this result.  The factor 2$H^2$ + $\dot{H}$ evaluates to 
\begin{equation}
2H^2 + \dot{H} = \frac{\kappa^2}{12b_{o}F(y_2)}\left(\rho_{1nr} + e^{-4A_{o}(y_2)}\rho_{2nr}\right)
\end{equation}
where
\begin{equation}
F(y) = \int_{0}^{y}dy'e^{-2A_{o}(y')}
\end{equation}
Thus $\delta$$\phi$($y_i$) depends only on the non-relativistic energy density.  The integrated bulk equations for $\delta$$\phi$(y) also gives a relations involving $\delta$$\phi$($y_i$):
\begin{eqnarray}\nonumber
\int_{0}^{1}dy\frac{\delta b}{b_o} &=& \sum_{i}(-1)^{i}\frac{\delta\phi(y_i)}{\phi_o(y_i)}+ \int_{0}^{1}\frac{dy}{A_{o}''}b_{o}^{2}\left(2A_{o}'e^{4A_o}F(y) + e^{2A_o}\right)\left(2H^2 + \dot{H}\right)\\&-& \int_{0}^{1}\frac{dy}{A_{o}''}\frac{\kappa^2}{6}b_{o}A_{o}'e^{4A_o}\rho_{1nr} 
\end{eqnarray}
where $\phi$($y_i$) is given by Eq.(34).  The left hand side of Eq.(37) is just the fractional change of the invariant distance between the branes.  Thus the Randall-Sundrum model accommodates a static solution (to $\mathnormal{O}$($\rho$)) with $\rho_{nr}$ $\neq$ 0 by allowing the distance between the branes to change by an amount proportional to $\rho_{nr}$.
\vspace{.4cm}

This work was supported in part by National Science Foundation Grant PHY-0101015 and in part by the Natural Science and Engineering Research Council of Canada.

\end{document}